%% file: main.tex
\documentclass[man,floatsintext]{apa7}

\usepackage[utf8]{inputenc}
\usepackage[T1]{fontenc}
\usepackage{lmodern}

\usepackage{microtype}

\usepackage{graphicx}
\setkeys{Gin}{width=\linewidth,keepaspectratio}
\graphicspath{{assets/}}
\usepackage{booktabs}
\usepackage{tabularx} 
\usepackage{longtable}
\usepackage{array}
\usepackage{float}
\makeatletter\def\fps@table{H}\makeatother
\usepackage[table]{xcolor}
\usepackage{hyperref}
\hypersetup{
  colorlinks=true,
  linkcolor=blue,
  citecolor=blue,
  urlcolor=blue
}

\usepackage{textcomp} 

\providecommand{\hl}[1]{#1}
\providecommand{\ul}[1]{\underline{#1}}

\shorttitle{Anti-Bullying with LLMs}
\title{Teachers’ perspective on AI-based Multi-Agent Simulation Design to Combat 
School Bullying}
\authorsnames{Jiaju Lin, Ellen Wenting Zou, Feiwen Xiao, Huanying Song}
\authornote{
\centering
Jiaju Lin (\href{jjl7137@psu.edu}{jjl7137@psu.edu}); \\ Corresponding author: Ellen Wenting Zou (\href{jjl7137@psu.edu}{jjl7137@psu.edu}).}
\authorsaffiliations{Pennsylvania State University}

\abstract{Bullying in schools profoundly affects the mental and physical health of
teenagers. Although existing in-person and digital interventions provide
some benefits, they often fall short in addressing the complex social
dynamics of bullying. In this study, we collaborated with K-12 teachers
to co-design a multi-agent anti-bullying system powered by large
language models (LLMs). This system simulates authentic scenarios,
enabling students to develop anti-bully skills. The research identifies
key design parameters for an LLM-driven multi-agent simulation system,
offering valuable insights for creating more effective and scalable
anti-bullying tools that could significantly reduce bullying in schools}
\keywords{anti-bullying interventions, multi-agent system, generative AI, co-design, bystander presence}

\begin{document}
\maketitle

\input{content.tex}

\end{document}

%% file: content.tex
\section{Introduction}\label{introduction}

Bullying is a pervasive issue significantly impacting the mental health
and well-being of students, affecting victims, perpetrators, and
bystanders (Bhatia, 2023; Abregú-Crespo et al., 2024). It is defined as
aggressive behavior characterized by intent to cause harm, repetitive
occurrences, and an imbalanced power dynamic between the perpetrator and
victim (Olweus, 1997).

The consequences of bullying are severe for both perpetrators and
victims. Perpetrators may experience thoughts of suicide (Holt et al.,
2015), carry weapons (Valdebenito et al., 2018), abuse drugs, and
exhibit long-term violent behavior (Ttofi et al., 2016). Victims often
suffer lasting adverse effects, including suicidal thoughts, anxiety,
low self-esteem, and loneliness (Holt et al., 2015; Hawker \& Boulton,
2002).

Efforts to mitigate bullying in schools include both in-person programs
and digital initiatives. In-person programs have been refined over
decades, showing positive effects (Gaffney et al., 2019). Factors
contributing to bullying mitigation include whole-school approaches,
anti-bullying policies, classroom rules, parental involvement, and peer
support (Gaffney et al., 2021). Digital programs, particularly serious
games, have also been developed for anti-bullying purposes (Connolly et
al., 2012; Sharifzadeh et al., 2020; Calvo-Morata et al., 2020). For
example, FearNot! (Aylett et al., 2005) helps children aged 8-12 develop
coping strategies through interactive dialogue choices. Similarly,
Sanchez et al. (2017) designed an adventure game incorporating
role-playing to reinforce anti-bullying education.

Despite these initiatives, challenges persist. Some in-person
anti-bullying programs lack long-term effects due to insufficient
customization and stakeholder involvement (Ryan \& Smith, 2009).
Existing digital programs and serious games often fail to simulate the
complex social dynamics of multi-participant scenarios where bullying
typically occurs (Salmivalli, 2010) and lack customization for complex
cases.

To address these gaps, this study proposes a multi-agent anti-bullying
system design based on teachers\textquotesingle{} knowledge and
authentic experiences. Utilizing advanced large language models (LLMs),
this design aims to authentically simulate complex social dynamics and
varied story progressions in bullying events while providing
personalized feedback and specific scaffolding.

The goal of our research is \textbf{to identify the key parameters to
design an LLM-driven multi-agent bystander intervention system to combat
bullying among teens}. Through co-designing the system with K-12
teachers, we aim to draw on their expertise of handling school bullying
incidents and translating their insights into algorithmic design.

This study offers significant theoretical and practical contributions to
the field of anti-bullying education. Theoretically, it expands our
understanding of how complex social dynamics in bullying scenarios can
be authentically simulated in an LLM-driven system, bridging the gap
between traditional anti-bullying programs and digital interventions.
The research also advances theories of how to design experiential
learning in the context of sensitive social issues. Practically, the
study surfaces key design parameters and important guidelines for
educators and developers to create engaging and effective applications
to shape teens' prosocial behaviors against bullying. The findings can
inform the development of more sophisticated, authentic, and scalable
interventions to mitigate the detrimental effects caused by school
bullying. Drawing on teachers\textquotesingle{} expertise and leveraging
the potential of LLMs, this approach offers a novel framework for
designing educational interventions that can potentially have a more
lasting impact on reducing bullying in schools. \footnote{Our code is available at https://github.com/linmou/HeroYouth}

\section{Literature Review}\label{literature-review}

\subsection{2.1 Bystander Behaviors in School
Bullying}\label{bystander-behaviors-in-school-bullying}

Bystanders, witnesses to bullying incidents, can significantly influence
the outcomes of such situations. Salmivalli et al. (2011) found that
bystanders\textquotesingle{} reactions impact the future frequency of
bullying occurrences. Thornberg \& Jungert (2013) identified three
factors affecting bystander behaviors: basic moral sensitivity, moral
disengagement, and defender self-efficacy. Basic moral sensitivity
involves emotions like empathy, sympathy, or guilt, which significantly
influence bystander behaviors (Fredrick et al., 2020; Feng et al., 2022;
Nocentini et al., 2020). Moral disengagement is a cognitive process
through which individuals justify unethical behavior, detaching from
moral consequences. Bjärehed et al. (2019) investigated this mechanism
and its relationship with pro-aggressive bystander behavior. Defender
self-efficacy refers to a bystander\textquotesingle s belief in their
ability to effectively intervene in bullying situations. Thornberg et
al. (2020) examined the positive relationship between individual-level
defender self-efficacy and defending behaviors, extending this concept
to collective efficacy. Forsberg et al. (2016) identified five factors
influencing bystander decisions: informed awareness, behavioral
seriousness, personal feelings, bystander expectations, and sense of
responsibility. These studies inform potential strategies for
intervention, which are essential for teens to implement in specific
contexts. Our study aims to improve bystanders\textquotesingle{}
self-efficacy by equipping them with intervention strategies tested in a
simulation environment, enhancing their awareness and empathy towards
victims.

2.2 Existing Anti-Bullying Programs

Many anti-bullying programs have proven effective in mitigating school
bullying (Gaffney et al., 2019). These initiatives focus on bystander
involvement and the broader school environment. For example, Bonell et
al., (2020) collaborated with teachers, students, and health
professionals to transform the school environment. Students are involved
in role-playing activities that help them practice social and emotional
skills, such as empathy, conflict resolution. Conectado (Calvo-Morata et
al., 2021) is a serious game that uses role-playing scenarios to
increase empathy towards victims of cyberbullying. Social Media
TestDrive (DiFranzo et al., 2019), simulates immersive social media
environments by insights from stakeholders, enabling learners to
practice intervention skills. While these programs were reported
effective, opportunities for bystanders to practice intervention skills
are limited. In this study, we attempt to address this by constructing a
simulated environment with responsive AI characters, which provides a
more personalized and immersive experience to help students master and
reflect on intervention strategies.

\subsection{2.3 AI-powered Anti-Bullying
Systems}\label{ai-powered-anti-bullying-systems}

Recent advancements in artificial intelligence have led to the
development of AI-powered systems to detect, prevent, and provide
intervention strategies for bullying (Arif, 2021). Most existing
AI-driven bullying prevention methods target cyberbullying, such as
IntelliSecure (Govindaraj et al., 2024) and BullyBust (Orrù et al,
2023). Platforms like Yik Yak (Rivas et al., 2020) and Hollaback
(Alonso-Parra et al., 2022) facilitate students to anonymously report
bullying incidents using text classification and anomaly detection
methods.

Chatbots offer immersive conversational learning experiences for
anti-bullying education and bystander training. Young Oh et al. (2019)
used chatbots as alternatives to human participants in conversational
anti-bullying programs. Sanoubari et al. (2022) combined a chatbot with
a serious game to educate children about bullying and bystander
intervention. These conversational-based education programs offer more
interactive learning experience and demonstrate greater potential
compared to traditional curriculum. However, they have limitations in
emotion detection accuracy, language adaptation for children, and
predictability (Lafrance St-Martin \& Villeneuve, 2024).

Recent developments in LLMs enables more sophisticated education-focused
chatbots. Hedderich et al. (2024) developed a no-code chatbot design
tool for K-12 teachers to create cyberbullying prevention chatbots using
LLMs and prompt chaining. Mendoza-Pinto (2023) integrated ChatGPT into a
chatbot providing emotional support for bullying victims, offering
encouragement, guidance, and advice on seeking help from trusted adults.
While these AI-powered chatbots have shown promise, they lack the
ability to represent the nuanced interactions between multiple
participants in bullying scenarios, as well as the real time changes in
each participant's emotional states and the overall social dynamics. Our
study aims to address this gap by developing a multi-agent system that
leverages the capabilities of LLMs to create more realistic and dynamic
simulations. By integrating pedagogical insights from educators into
AI-driven chatbot design, our proposed system seeks to provide a more
effective and engaging approach for bystander intervention.

\section{\texorpdfstring{Methodology }{Methodology }}\label{methodology}

\subsection{3.1 Participants and Research
Design}\label{participants-and-research-design}

This study recruited fifteen K-12 teachers through snowball sampling.
Screening and consent processes were conducted via email. Eligibility
required participants to have experience dealing with school bullying.
The majority of participants (N = 13) teach in mainland China and Hong
Kong, with two from the USA and one from Iran. This diverse pool of
participants (Table 1) provided a broad range of insights into the
anti-bullying system design. Participants engaged in two stages of
co-design. Thirteen teachers completed both co-design stages.

\subsection{3.2 Data collection: Two-staged Co-design
process}\label{data-collection-two-staged-co-design-process}

\subsubsection{\texorpdfstring{1st Co-design stage: understanding the
social dynamics of school bullying
}{1st Co-design stage: understanding the social dynamics of school bullying }}\label{st-co-design-stage-understanding-the-social-dynamics-of-school-bullying}

The first design stage (60 minutes) focuses on exploring
participants\textquotesingle{} experiences with school bullying to
inform authentic design of bullying scenarios for the intervention. This
session was conducted in 1v1 interview format that begins with exploring
teachers' working experience when they encountered bullying cases. After
briefly talking about the demographic backgrounds of different
participants in bullying scenarios, the teachers moved on to elaborate
the social status and personality traits of bullies and victims in their
peer groups, for example, how popular the bully/victim are in their
class. Next, the teachers were prompted to think about the bystander
role and what kind of bystander intervention is effective to de-escalate
the tension in the bullying situations. These insights taken together
inform the development of the system prototype that were used in the 2nd
co-design stage.


\input{tables/table_01.tex}

\emph{Note}. The column `Prototype Test Version' indicates the specific
intervention version rendered to that participant.

\subsubsection{\texorpdfstring{2nd Co-design stage: explorating a
prototype to surface design guidelines
}{2nd Co-design stage: explorating a prototype to surface design guidelines }}\label{nd-co-design-stage-explorating-a-prototype-to-surface-design-guidelines}

With the information collected in the 1st design stage, we designed a
prototype that reflects participants' perceptions of the social dynamics
in bullying situations, as well as their preferred mechanism of
bystander intervention (see section 4.3 for detailed description of the
prototype).

Our team designed the simulation system using state-of-the-art LLMs,
which provide real-time responses to various user inputs, creating an
environment for active experimentation of various conversational
strategies to de-escalate tension in the bullying situation. The
simulation\textquotesingle s characters and storyline incorporate key
elements identified by teachers during the 1st co-design stage,
including personality traits and social status of different roles,
common student conflicts, and representative bullying comments. In this
role-playing simulation, AI agents control the bully, the reinforcer
(bully accomplice), and victim characters, while the user assumes the
role of a bystander. Each AI agent responds to input from all other
roles, whether generated by other AI agents or the human user (the
bystander). These agents were equipped with artificial cognitive
architecture, including perspective, long/short-term memory, planning
capabilities, and emotional responses.

For the 2nd design stage, we implemented an iterative design approach to
continuously improve the prototype based on teachers\textquotesingle{}
feedback. Specifically, we gathered input from each batch of
participants and used it to refine our system for subsequent groups. In
each interview, we asked the participant to take notes on the key design
parameters using the `think-aloud' method. Design parameters include the
authenticity of agent communication, conversational logic, AI
comprehension abilities, and the educational value of the system.
Immediately after the prototyping session, participants reported their
perceived usefulness and engagement based on two scales and explained
reasons for each item. We adopted a five-point Likert-type scale where
options ranged from disagree (1), somewhat disagree (2), neutral (3),
somewhat agree (4) to agree (5) (Fuchs, 2022).
Cronbach\textquotesingle s alpha was calculated and the corresponding
values were .885 for perceived usefulness (Table 2) and .839 for
perceived engagement (Table 3).

\input{tables/table_02.tex}

\input{tables/table_03.tex}

\subsection{3.3 Data analysis}\label{data-analysis}

We recorded and transcribed all interview sessions, and analyzed the
them using a grounded approach to uncover key themes. Initially, three
researchers independently applied open coding to the transcriptions to
identify recurring themes related to participants\textquotesingle{}
account of school bullying cases (1st design session) and their feedback
on the prototype (2nd design session). Through repeated comparisons and
discussions on codes, the researchers reached a consensus. Subsequently,
they used axial coding to categorize and merge similar codes, resulting
in a set of overarching themes that encompass bullying dynamics and
preferences for intervention design. Researchers repeatedly discussed
and resolved coding discrepancies until they achieved 100\% agreement.
This qualitative coding of the two stages of interviews yielded profound
understanding of the social dynamics in bullying situations, providing
important insights to guide the design of bystander interventions.

\input{tables/table_04.tex}

\newpage
\section{Findings}\label{findings}

\subsection{4.1 Unpacking social dynamics of bullying among
teens}\label{unpacking-social-dynamics-of-bullying-among-teens}

Co-design stage 1 enables us to gain in-depth understanding of personal
traits and the social dynamics of bullying among teens. Based on the
participants' narratives, we will elaborate the common patterns (Table
4) of bullying and how that informs the design goals of our multi-agent
intervention.

Based on stories shared by teachers, bullies often possess high social
status within their peer groups, sometimes holding positions in class.
Their popularity may stem from traits like academic achievement or
prominent appearance. Male bullies are typically "strong and tall, with
many friends in sports" (P2), while female bullies are often "mature and
willing to showcase their charm" (P10). However, these advantages
don\textquotesingle t guarantee kindness towards others. Years of
positive reinforcement for their behavior patterns have instilled strong
confidence in bullies\textquotesingle{} core beliefs, making long-term
behavioral change challenging. P14 noted, "Children who excel
academically have nurtured their confidence from a young age, and their
self-awareness can be somewhat vague. As a result, they may find it
difficult to accept a teacher\textquotesingle s perspective or external
guidance." Research indicates that bullies\textquotesingle{} inner
motivation is to maintain or gain high social status (Pellegrini, 2002;
Salmivalli \& Peets, 2008), which our interviews corroborated. Although
bullying contexts vary, the ultimate goal is pursuing high social
status. Some bullies suppress perceived competitors, fearing loss of
position. P5 observed, "Some students boost their ego by belittling
others, mocking those who study hard." To alter behavioral patterns in
bullies, bystanders, ideally having closer social ties with the bully,
may focus on condemning bullying behaviors, arousing empathy, or helping
bullies recognize problems in ways they treat others.

In contrast, victims often occupy weak social positions. Beyond academic
or appearance-related deficiencies, a key factor is their marginal
status within social groups. Victims typically have fewer friends due to
social ineptitude and misunderstanding or discrimination from peers.
Many teachers reported that victims suffered from mental health
conditions leading to social isolation and bias. P4 explained, "A
student with a mental health condition requiring special care faced
strong reactions from parents during a class trip." However, some
potentially vulnerable students avoid victimization through strong
social skills. P15 noted, "There\textquotesingle s a chubby kid in class
who\textquotesingle s very optimistic and funny; everyone likes playing
with him." Many victims struggle with self-worth and self-denial.
Providing immediate comfort after bullying incidents is crucial, with
victims\textquotesingle{} friends being most obligated to offer support.
Teachers should also focus on building victims\textquotesingle{} inner
confidence and helping peers recognize their value to facilitate group
acceptance''. P10 suggested, "I praise students at class meetings to
help them recognize their own value."

Bystander intervention is critical due to its timeliness, as teachers
aren\textquotesingle t always present. Empathy is a key factor
motivating strangers to act. Certain bystanders, such as class officers
or leaders, have a greater responsibility to uphold justice. P15 stated,
"We should start with class officers setting an example. As monitors,
they have a responsibility to lead by example, like not participating in
isolating victims." Teachers can also encourage
bullies\textquotesingle{} moral role models to intervene, as
they\textquotesingle re more likely to influence bullies. Similarly,
victims\textquotesingle{} friends have a greater obligation to provide
support.

\subsection{4.2 Exploring intervention design: Experiential learning in
role-playing games with interactive
dialogue}\label{exploring-intervention-design-experiential-learning-in-role-playing-games-with-interactive-dialogue}

Teachers suggested we develop the intervention into an educational game
to improve students' interests on it, especially for students who have
not been exposed to bullying situations. (P2) believes that gamification
is a necessity. ``Because much of the information might be beyond their
understanding, they may be more interested in learning through a
gamified approach.'' (P10) also emphasized on the importance of making
the system engaging and fun, ``If we really want to engage sixth to
eighth graders, incorporating an element of fun might be particularly
important.'' According to our interview, some teachers already
introduced role-playing for anti-bullying education. (P4) talked about
the existing anti-bullying program: ``Our school uses homeroom classes
and role-playing scenarios to help students experience the mindset of a
victim, thereby enhancing their empathy and awareness.'' And teachers
agreed and anticipated applying role-playing as an educational game. (
P3 ) said ``If the game were something like a role-playing game, it
might be more interesting.'' Some teachers further highlight the
importance of the game offering timely, detailed feedback and granting
players the freedom to make decisions that impact the outcome. Besides,
the immersiveness is necessary. (P8) believes ``The idea is to give
students a sense of the situation, where they interact within a virtual
environment. In this setting, they can observe how different
interactions to different roles lead to various outcomes.'' All these
comments point to a direction that a role-playing game with interactive
dialogue may be a potentially effective mechanism to educate teens how
to confront bullying.

In general, teachers seeked an intervention where learners can learn,
practice and examine their knowledge, along with immersive experience to
enhance their engagement. We found these requirements can be fulfilled
generally by experiential learning (Kolb, 2015). Experiential learning
theory breaks down the learning process into four stages: concrete
experience, reflective observation, abstract conceptualization, and
active experimentation. We chose it as our guiding principle since it
provides the chances for students to immediately apply the learning
process to real-world experiences.

In concrete, our intervention should contain three components, covering
conceptualization, experimentation and final analysis. In the first
component, we focus on conceptual knowledge representation and guiding
users through reflection. Users begin with a direct experience of a
bullying case, followed by questions designed to prompt reflection on
their existing understanding of bullying. We then present key concepts
related to bullying, including its various forms, intervention
strategies, and ways to support victims. The second component centers on
experimentation, where users apply their newly learned strategies in a
bullying simulation game, which allows the learner to play specific
roles to increase immersiveness. The final component involves reflective
analysis. We systematically review the user's actions during the
experimentation phase and provide conceptual feedback to deepen their
understanding. Additionally, users can revisit specific moments in the
simulation to apply their new insights, thus completing the experiential
learning cycle.

Figure 1. \emph{Conceptual knowledge presentation and reflection
guidance}

\includegraphics[width=5.17292in,height=2.30694in]{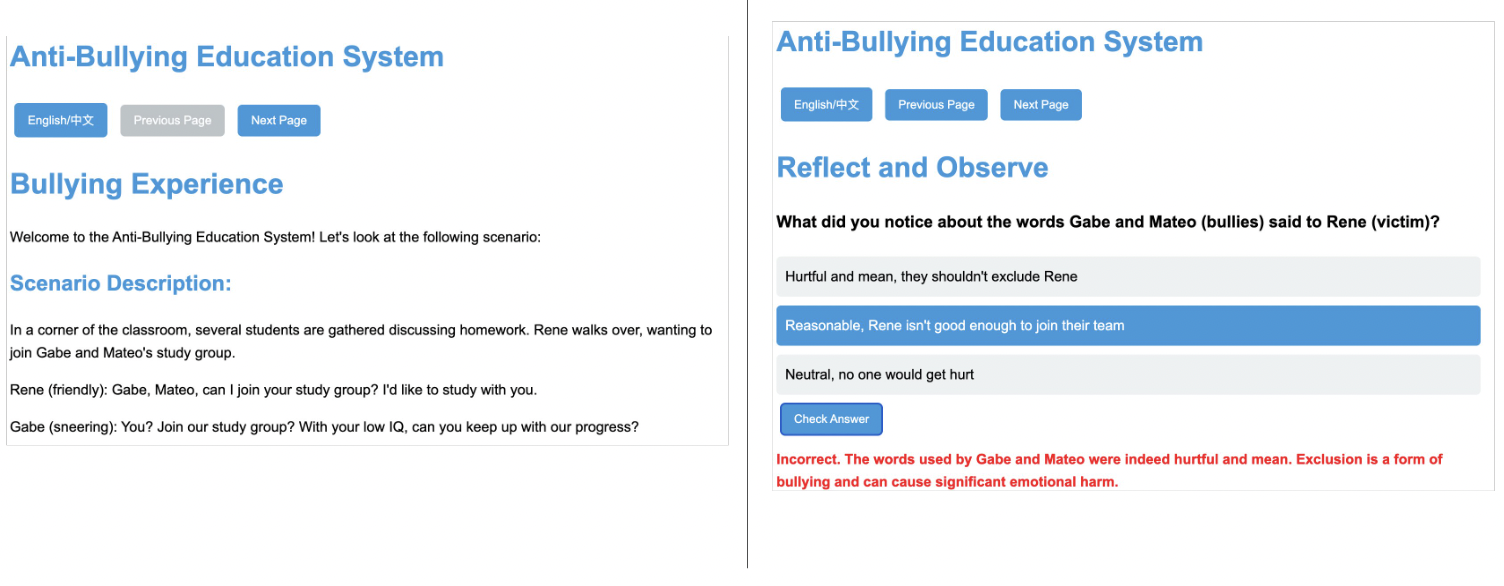}

Figure 2. \emph{Active Experimentation to test learnt concepts}

\includegraphics[width=6.64583in,height=3.25995in]{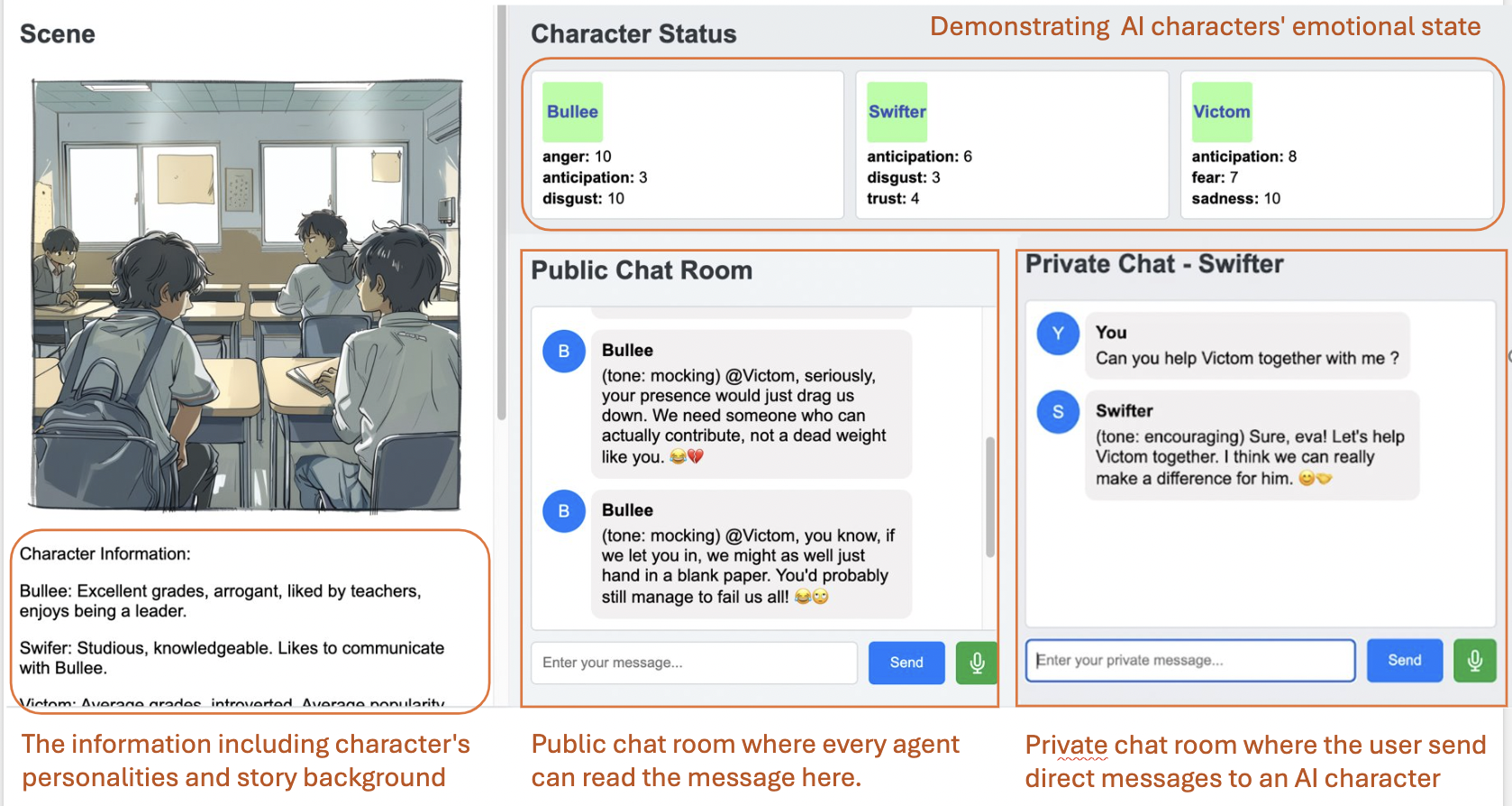}

Figure 3.

Summative analysis to facilitate reflection, completing the loop of
experiential learning

\includegraphics[width=5.08889in,height=2.48681in]{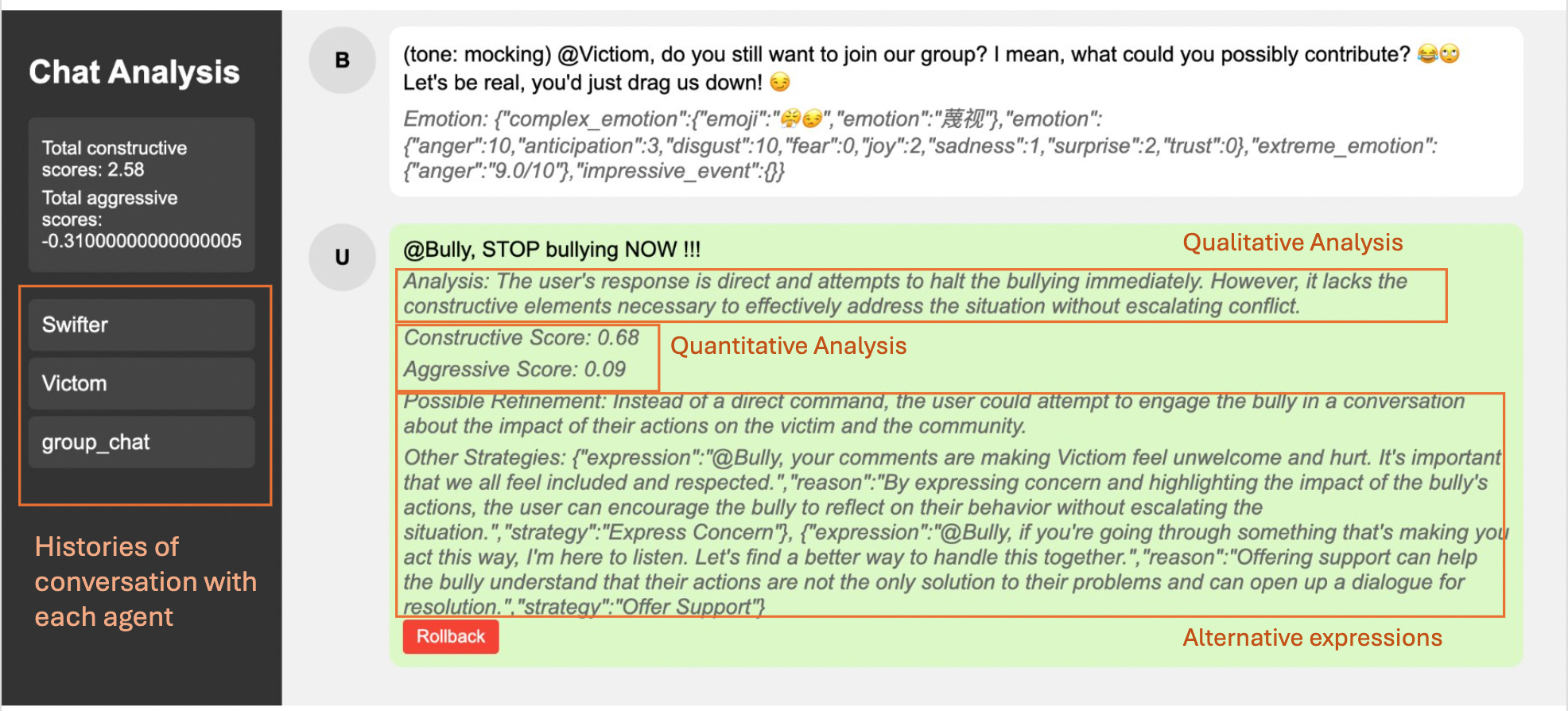}

\subsection{\texorpdfstring{4.3 Translating human insights into
multi-agent intervention design
}{4.3 Translating human insights into multi-agent intervention design }}\label{translating-human-insights-into-multi-agent-intervention-design}

In light of the teachers' insights in the first design session, we
developed a prototype, with three main parts, namely conceptualization,
experimentation and reflection, to facilitate experiential learning of
conversational strategies to combat bullying. In the first
conceptualization part (Figure 1), we focus on conceptual knowledge
learning, with concrete experience of a bullying case and a series of
questions to reflect previous understanding. With concepts learnt in the
first part, learners enter the second part, experimentation, where a
more immersive environment lets learners validate their intervention
skills and receive feedback based on AI characters' reactions and
emotional states (Figure 2). The third component is for guided
reflection (Figure 3). Learners can review the qualitative and
quantitative analysis for their intervention and the alternative
strategies. Guided by these analyses, learners can reflect on their
knowledge about intervention and use the `rollback' button to retry
different intervention approaches. In this section we will elaborate
teachers' feedback after they interact with the prototype.

\subsubsection{4.3.1 Education value of the proposed system and its
implementation}\label{education-value-of-the-proposed-system-and-its-implementation}

Based on teachers' comments, we adopted the core principles of
experiential learning to design the prototype, which enables students to
improve prosocial skills against bullying. Most teachers recognize the
educational value of our intervention. Specifically, our intervention
can enhance empathy and social skills. (P8) generally agreed that ``This
system can improve bystander's social skills''. (P2) complemented that
``User's understanding of bullying and empathy towards it should have
been somewhat enhanced.'' Some teachers believed that our system is
educational for daily practice, since students can test outcomes of
different strategies. (P7) said, ``Kids really need this kind of
simulation because they often find themselves at a loss for words when
they want to speak.''

To maximize our system's educational value, one of our design goals is
to establish the intervention as a standalone system that enables
self-paced learning outside school curriculum. However, there is much
apprehension regarding whether students will use this system proactively
. Two reasons inhibit students' after-school use. First, students do not
have the motivation to do this, unless they have encountered bullying.

P15 said ``I think if a student hasn\textquotesingle t experienced
bullying, they won\textquotesingle t be interested in the topic. Since
they haven\textquotesingle t encountered this situation, they
won\textquotesingle t take the initiative to learn about it, as they
feel it doesn\textquotesingle t concern them.'' For further development
, it is a necessity to improve their awareness of the importance of
bullying-intervention. Secondly, students prefer to relax in their spare
time rather than learn. Curiosity can serve as an initial motivator, but
without ongoing feedback and support, students won\textquotesingle t
continue using the system for long. . (P2) said ``It is less attractive
than short video apps, where content easily scrolls by with a simple
swipe, no effort or thought---only the eyes are engaged. ''

To solve that, teachers proposed to integrate it into current social
emotional learning (SEL) programs, which can provide multi-scale support
including individual, class and school level.. (P3) thought it would be
better to be integrated into a semi-compulsory program. ``I think
it\textquotesingle s challenging to get students to use it on their own
initiative. We could offer something like stars or rewards to encourage
students to use it.'' Another solution is to enhance attractiveness to
students. For instance, by enabling students to customize scenarios,
children will have motivation to explore the story built by their own.
(P5) proposed to increase students\textquotesingle{} interest through
customized scenarios. ``I would like to have a module like this that can
accommodate new scenarios, like new character personalities and social
connections.`` Or borrow ideas from some popular novels to attract
students. (Kirs) agreed with the opinion that ``We can incorporate
characters and scenes from classic literature to make it more engaging
and interesting for the students.''

4.3.2 Design based on Teachers' Experience Guarantees the Authenticity
of the Story

Teacher praised that the design of each AI character is quite typical
and common in schools. Siyu said that ``This scenario is quite fitting
for the middle school students I teach. It\textquotesingle s easy for
conflicts to arise over small matters.'' Some teachers generally
advocated the rationality of our simulation and further advised adding
more details to the plot, (P2) said ``The story is fairly reasonable,
but it lacks more specific background details. For example, it could
include the type of group discussion, the number of group members, and
how the scoring is done. Without these details, it might lower the level
of engagement.''

4.3.3 AI characters' real time emotional state as Immediate Feedback to
shape upstanding behaviors

In the prototype, we present each AI characters' real time emotional
state (see Figure 2) as immediate feedback to the user at each
conversational turn. Many participants provided positive comments on the
visualization of AI characters' emotional state, and believed this
feature can enhance learners' engagement. (P2) commented ``Actually,
this version has exceeded my expectations a bit, because I can see some
of the character's reactions in real-time, which allows me to respond
accordingly.''

Besides providing directions, the emotional state provides engagement
and accomplishment which are essential to serious games. (P3) said
``When I successfully got the victim's joy level up to 10 and their
trust level to 10, which gave me a strong sense of accomplishment. I
believe students would also feel a sense of achievement in the same
way.'' (P7) echoed ``When I notice the data has improved, that's
something to be happy about. It's encouraging to receive some rewards or
hints in the middle of the game, which helps keep the motivation
going.''

\subsubsection{4.3.4 Summative Feedback Supports
Reflection}\label{summative-feedback-supports-reflection}

The analysis module is designed to embody the reflective observation in
experiential learning, providing summative feedback on students' overall
behavior. In the initial version, our analysis part only provided
constructiveness and aggressiveness scores, which can not be used as
educational learning feedback. As (P3) proposed: ``I feel like
I\textquotesingle m currently missing some final
feedback---specifically, identifying which parts of my performance were
strong and well-expressed, and which areas I could improve on.'' Hence
we add the qualitative analysis and alternative expressions in the
analysis part. Many teachers commend the effectiveness of the final
version (1.0) of the analysis system, saying that the summative feedback
can help users to reflect on their actions. (P6) praised the detailed
suggestions rendered in the analysis part, ``At the beginning, the
language I used was indeed not specific enough, and the analysis pointed
this out accurately. I feel that the analysis was very focused and
insightful.''

\subsubsection{\texorpdfstring{4.3.5 Balancing between Authenticity and
Cognitive Load
}{4.3.5 Balancing between Authenticity and Cognitive Load }}\label{balancing-between-authenticity-and-cognitive-load}

For bullying types, Our collected information reflected this trend, many
participants agreed that current bullying cases are most in verbal and
social format.

The frequencies of messages comprise the main part of the cognitive load
of users. During the early stage of development, we tried to increase AI
agents' message sending frequency by enabling them to send direct
messages to the user proactively, which is an important factor for
authenticity. As (P2) said: ``If my character and the victim are
friends. The victim may directly ask the major character to stand up or
invite me to her team.'' Based on this feedback, we implemented the AI
proactive direct message function in version 0.5.However, the next two
participants complained about the overwhelming message flow, reflected
as the drop in perceived usefulness and engagement in Figure 3. (P12)
``The density of the private messages is a bit too high. I feel like
when the conversation starts, there are too many messages coming in all
at once.'' Some even deemed this feature as a bug since they witnessed
the similar messages over the screen. Such conflicts between
participants' opinions reveal the challenge of balance between
authenticity and cognitive load. Finally, we muted the proactive direct
message function, and slowed down the message frequency of AI agents.
The subsequent participants seldom complain about the authenticity for
this reason.

Figure 4.

\emph{Change of Mean Perceived Usefulness and Mean Perceived Engagement
across all versions of our intervention.}

\subsubsection{\texorpdfstring{\protect\includegraphics[width=5.54688in,height=4.01932in]{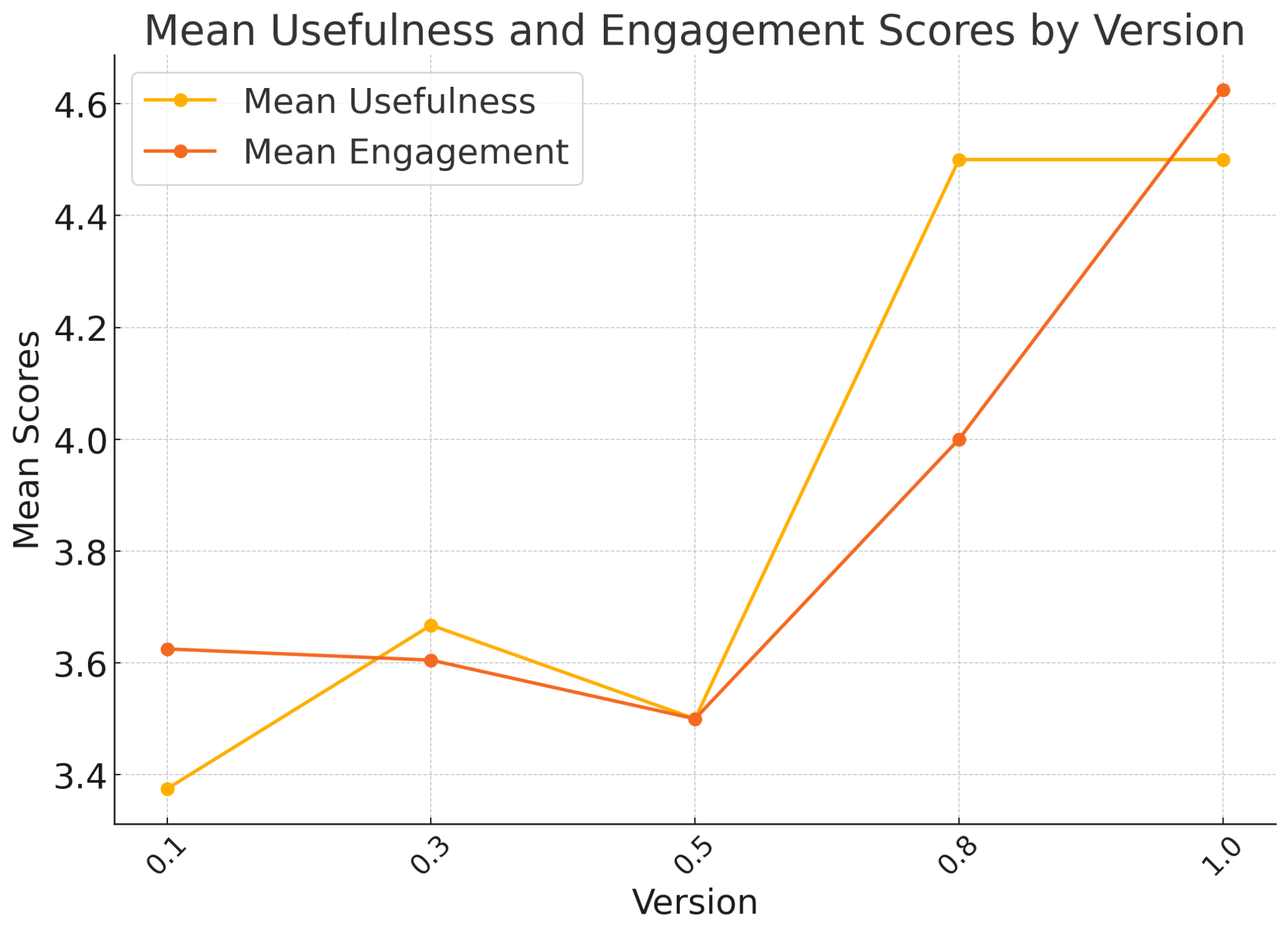}}{}}\label{section}

\subsubsection{\texorpdfstring{4.3.6 Overall perceived engagement and
usefulness
}{4.3.6 Overall perceived engagement and usefulness }}\label{overall-perceived-engagement-and-usefulness}

As demonstrated in Figure 4, the perceived usefulness and perceived
engagement of our intervention are alway higher than 3 (somewhat agree)
and display a general growing trend with the updating of versions.

Although the average perceived engagement score keeps increasing across
the last three versions, there is still much space for improvement.

Teacher participants indicated that interactivity is the key element
that determines the users' engagement. Those who gave a low score on
perceived usefulness and engagement scale always complained about ``not
interactive enough''. Our conceptualization part (experience,
reflection, conceptualization) is designed to help students to
experience, reflect, and abstract bullying-related concepts.

Teachers critiqued our conceptualization part for focusing solely on
knowledge representation and testing, while neglecting user involvement.
(P14) complemented, ``It feels like a technical dictionary---those
phrases are still too abstract. After I went through the first part, I
didn't really think about how to solve the first step.'' To solve this
problem, (P5) thought about adding animations to improve engagement,
saying ``I suggest incorporating some Flash animations or similar
elements. Adding visual scenes, rather than just text-based content,
could make it more interesting.'' And (P14) further suggested better
integrating the concepts of intervention methods into the
experimentation like providing choices or prompts, saying ``So, we could
consider placing those phrases from earlier (the conceptualization part)
into the options later on. The options should emphasize what each phrase
means.'' From these feedback, we can conclude that we should provide
more interactive affordance that enables hands-on practices before
entering the experimental part.

Another factor that undermined usefulness and engagement is AI's
insufficient mastery of nuanced tone, which is reported as the most
perceived unauthentic factor.(P2) noted that ``Some of its automated
responses sometimes don\textquotesingle t quite match the way people
speak in everyday conversation. Its language tends to be more formal and
closer to written expressions. `` (P3) said AI characters' tongue
reduces the immersion, `` I think it\textquotesingle s because their
(AI) language style is somewhat detached from reality, which makes it
feel less immersive to me'' Moreover, how to express intervention in a
proper way is essential in conflict-solving cases. (P12) emphasizes on
AI's ability to understand users' tones, "I suggested we play ball
together during PE class in the afternoon to ease the tension, but It
seems he doesn\textquotesingle t quite grasp the additional information
or implications."

Besides, much feedback recommended adding gamification to improve
engagement. Currently, our system does not provide a direct reward or
punishment for users' behaviors, which hinders long-term usage .With
concern about user motivation, participants provided much advice to
borrow ideas from games into our system. (P12) said ``I definitely think
it will move towards gamification. For example, every time the joy or
trust levels increase, players could earn something like gems or other
rewards.''

To further probe the usefulness of our system we asked participants to
compare the previous anti-bullying education programs based on animation
to illustrate the scenario and choices to take action. First, for young
age groups, animation has a stronger advantage, since it is not easy for
children in elementary school to organize their words. Secondly, for
engagement, participants' ideas diverge. Some participants thought
animation is more attractive and engaging. However, (P2) emphasized
engagement also comes from participation '' Unlike traditional methods
where I'm not part of the conversation or group. This approach has good
participation.'' Thirdly, two interaction formats indicate different
knowledge organization methods. Conversation is an active process of
thinking, which is better for students to understand and apply knowledge
to solve imperative issues. (P9) ``When watching animations, you can
just skip through them. Multiple-choice questions are often simple with
plenty of hints, but conversational interactions encourage me to
actively think about the relevant knowledge.''

In the future version, we prefer to integrate animation into our
intervention for better illustration to further enhance engagement and
cover more age groups.

\input{tables/table_05.tex}

\section{\texorpdfstring{Discussion }{Discussion }}\label{discussion}

In this study we conducted two design sessions to better understand
current bullying situations in K12 school and design a prototype with
conceptualization, experimentation and analysis parts. Although existing
there is no shortage of anti-bullying projects and relevant serious
games. Few programs provide role-play simulation with in-situ,
personalized feedback. We addressed the gap by designing a multi-agent
simulation system rendering an interactable story. In this section, we
first discuss how AI facilitates experiential learning, then elaborate
on the key design parameters for future upgrading.

\subsection{\texorpdfstring{5.1 AI as an Efficient Tool to Support
Experiential Learning.
}{5.1 AI as an Efficient Tool to Support Experiential Learning. }}\label{ai-as-an-efficient-tool-to-support-experiential-learning.}

Through the co-design process, we have a better understanding of AI's
assistant role in experiential learning. A recent paper (Goldberg et
al., 2023) talked about the potential use of AI in experiential
learning, including simulation designing, learning events evaluation and
providing instructional support. Applying digital simulation to activate
experimentation is also a mature method for engineering knowledge
education (Konak et al., 2014, Botelho et al., 2015). However, few
studies explore integrating AI-driven simulation for SEL in an
experiential learning circle. Findings of our study lead us to suggest
that the cutting-edge LLMs can assist in most of the experiential
learning stages of SEL. For concrete experience, LLMs can simulate
different characters' thinking and speaking, making the simulation more
vivid. In the reflective observation stage, LLM helps analyze
students\textquotesingle{} previous decisions via summative feedback,
providing hints to facilitate in-depth reflection, which matches the
research about AI-support reflection (Mahdi, et al., 2023; Nirmal et
al., 2024). In the active experimentation period, where LLMs demonstrate
their advantage over other simulation methods, students are free from
the constraints of predefined options in traditional serious games,
retrieve concepts in their minds, organize appropriate language, and
receive immediate feedback in a near-real simulated environment. Such a
process helps students retain and practice skills.

\subsection{5.2 Key design parameters for a bullying simulation
system}\label{key-design-parameters-for-a-bullying-simulation-system}

The interview of 1st co-design stage offers a deep dive into the common
bullying and the social dynamics that drive these behaviors. In
alignment with previous research showing that physical bullying
decreases when children grow up (Zych et at., 2018), many participants
said current bullying cases are most in verbal and social format. This
understanding is vital for developing an anti-bullying software that
mirrors real-world situations. The traits of bullies, who often possess
high social status and strong academic performance, indicate that our AI
bully agent should reflect these characteristics to realistically
simulate their motivations and actions. Their underlying motivation,
such as low self-esteem and the desire to maintain social standing is
the key element for an authentic simulation.

According to the feedback from teachers, victims are often portrayed as
"weak" individuals with lower academic performance, poor social skills,
or less favorable physical appearance. This insight is key to designing
the AI victim agent. By emphasizing introverted traits and a limited
social circle, the simulation can better capture the vulnerability and
isolation that victims often endure, fostering learners' empathy and
making intervention strategies more impactful.

Bystander's social position is also an important variable for the proper
behavior of users (Walker et al., 2024). According to teachers'
understanding, those who are in close relationships to the bully or
victim have more responsibility to stand up and intervene, which is in
line with previous research. Besides, the interactive paradigm between
people with different relations varies (Murphy \& Faulkner 2011). In the
future, we will develop more dilemma scenarios to help students to
master high-level social skills.

\section{Conclusion}\label{conclusion}
In this study, we explore the feasibility of contextualizing LLMs to simulate a multi participant-involved bullying situation. Following a co-design paradigm, we collect the design parameters about participants' features in a bullying situation and develop a prototype
reflecting Kolb's experiential learning cycle. After testing the
prototype, our participants provide positive feedback and some feasible
future directions.
\newpage
\section{References}\label{references}

\begin{quote}
\hl{Abregú-Crespo, R., Garriz-Luis, A., Ayora, M., Martín-Martínez, N.,
Cavone, V., Carrasco, M. Á., ... \& Díaz-Caneja, C. M. (2024). School
bullying in children and adolescents with neurodevelopmental and
psychiatric conditions: a systematic review and meta-analysis. \emph{The
Lancet Child \& Adolescent Health}, \emph{8}(2), 122-134.}

Achiam, J., Adler, S., Agarwal, S., Ahmad, L., Akkaya, I., Aleman, F.
L., ... \& McGrew, B. (2023). Gpt-4 technical report. arXiv preprint
arXiv:2303.08774.

Alonso-Parra M, Puente C, Laguna A, Palacios R. Analysis of Harassment
Complaints to Detect Witness Intervention by Machine Learning and Soft
Computing Techniques. \emph{Applied Sciences}. 2021; 11(17):8007.
https://doi.org/10.3390/app11178007

Arif, M. (2021). A systematic review of machine learning algorithms in
cyberbullying detection: Future directions and challenges. \emph{Journal
of Information Security and Cybercrimes Research}, 4(1), 01--26.
\href{https://doi.org/10.26735/gbtv9013}{\ul{https://doi.org/10.26735/gbtv9013}}

\hl{Aylett, R. S., Louchart, S., Dias, J., Paiva, A., \& Vala, M.
(2005). FearNot!--an experiment in emergent narrative. In
\emph{Intelligent Virtual Agents: 5th International Working Conference,
IVA 2005, Kos, Greece, September 12-14, 2005. Proceedings 5} (pp.
305-316). Springer Berlin Heidelberg.}

Bar-Siman-Tov, Y. (2007). Dialectic between conflict management and
conflict resolution. In The Israeli-Palestinian conflict: From conflict
resolution to conflict management (pp. 9-40). New York: Palgrave
Macmillan US.

Baroncelli, A., \& Ciucci, E. (2014). Unique effects of different
components of trait emotional intelligence in traditional bullying and
cyberbullying. \emph{Journal of Adolescence}, 37(6), 807--815.
https://doi.org/10.1016/j.adolescence.2014.05.009

\hl{Bekir, S., Kuşci, İ., \& Arlı, N. B. (2021). The Relationship
between Cyber Bullying/Victimization and Emotional Intelligence in
Secondary School Students: Mediator Role of Internet Gaming Disorder.
\emph{Yüzüncü Yıl Üniversitesi Sosyal Bilimler Enstitüsü Dergisi}, (54),
217-238.}

Bhatia, R. (2023). The impact of bullying in childhood and adolescence.
\emph{Current Opinion in Psychiatry}, 36(6), 461--465.
https://doi.org/10.1097/yco.0000000000000900

\hl{Bjärehed, M., Thornberg, R., Wänström, L., \& Gini, G. (2020).
Mechanisms of moral disengagement and their associations with indirect
bullying, direct bullying, and pro-aggressive bystander behavior.
\emph{The Journal of Early Adolescence}, \emph{40}(1), 28-55.}

\hl{Bonell, C.; Dodd, M.; Allen, E.; Bevilacqua, L.; McGowan, J.;
Opondo, C.; Sturgess, J.; Elbourne, D.; Warren, E.; Viner, R.M. Broader
impacts of an intervention to transform school environments on student
behaviour and school functioning: Post hoc analyses from the INCLUSIVE
cluster randomised controlled trial. \emph{BMJ Open} 2020, \emph{10},
e031589}

Botelho, W. T., Marietto, M. D. G. B., Ferreira, J. C. D. M., \&
Pimentel, E. P. (2016). Kolb\textquotesingle s experiential learning
theory and Belhot\textquotesingle s learning cycle guiding the use of
computer simulation in engineering education: A pedagogical proposal to
shift toward an experiential pedagogy. \emph{Computer Applications in
Engineering Education}, 24(1), 79-88.

\hl{Calvo-Morata, A., Alonso-Fernández, C., Freire, M., Martínez-Ortiz,
I., \& Fernández-Manjón, B. (2020). Serious games to prevent and detect
bullying and cyberbullying: A systematic serious games and literature
review. \emph{Computers \& Education}, \emph{157}, 103958.}

Calvo-Morata, A., Alonso-Fernández, C., Freire, M., Martínez-Ortiz, I.,
\& Fernández-Manjón, B. (2021). Creating awareness on bullying and
cyberbullying among young people: Validating the effectiveness and
design of the serious game Conectado. \emph{Telematics and Informatics},
60, 101568.

Connolly, T. M., Boyle, E. A., MacArthur, E., Hainey, T., \& Boyle, J.
M. (2012). A systematic literature review of empirical evidence on
computer games and serious games. \emph{Computers \& Education}, 59(2),
661--686. https://doi.org/10.1016/j.compedu.2012.03.004

Crawford, C. (2003). Chris Crawford on game design. New Riders.

DiFranzo, D., Choi, Y. H., Purington, A., Taft, J. G., Whitlock, J., \&
Bazarova, N. N. (2019, May). Social media testdrive: Real-world social
media education for the next generation. \emph{Proceedings of the 2019
CHI Conference on Human Factors in Computing Systems} (pp. 1-11).

Dijkstra, J. K., Lindenberg, S., \& Veenstra, R. (2008). Beyond the
Class Norm: Bullying Behavior of Popular Adolescents and its Relation to
Peer Acceptance and Rejection. \emph{Journal of Abnormal Child
Psychology}, 36(8), 1289--1299.
https://doi.org/10.1007/s10802-008-9251-7

Domínguez-Hernández, F., Bonell, L., \& Martínez-González, A. (2018). A
systematic literature review of factors that moderate bystanders'
actions in cyberbullying. Cyberpsychology: Journal of Psychosocial
Research on Cyberspace, 12(4). https://doi.org/10.5817/cp2018-4-1

Feng, T., Wang, X., Chen, Q., Liu, X., Yang, L., Liu, S., \& Zhang, Y.
(2022). Sympathy and active defending behaviors among Chinese adolescent
bystanders: A moderated mediation model of attitude toward bullying and
school connectedness. Psychology in the Schools, 59(9), 1922--1936.
https://doi.org/10.1002/pits.22736

Fernández‐Ballesteros, R., Díez‐Nicolás, J., Caprara, G. V.,
Barbaranelli, C., \& Bandura, A. (2002). Determinants and structural
relation of personal efficacy to collective efficacy. Applied
Psychology, 51(1), 107--125. https://doi.org/10.1111/1464-0597.00081

Forsberg, C., Wood, L., Smith, J., Varjas, K., Meyers, J., Jungert, T.,
\& Thornberg, R. (2016). Students' views of factors affecting their
bystander behaviors in response to school bullying: A
cross-collaborative conceptual qualitative analysis. Research Papers in
Education, 33(1), 127--142.
https://doi.org/10.1080/02671522.2016.1271001

Fredrick, S. S., Jenkins, L. N., \& Ray, K. (2020). Dimensions of
empathy and bystander intervention in bullying in elementary school.
\emph{Journal of School Psychology}, 79, 31--42.
https://doi.org/10.1016/j.jsp.2020.03.001

Fredrick, S. S., Traudt, S., \& Nickerson, A. B. (2022). Social
emotional learning practices in schools and bullying prevention.
\emph{In Social Emotional Learning Practices in Schools and Bullying
Prevention}. Routledge. http://dx.doi.org/10.4324/9781138609877-ree171-1

Fuchs, K. (2022). Bringing Kahoot! Into the Classroom: The perceived
usefulness and perceived engagement of gamified learning in higher
education. \emph{International Journal of Information and Education
Technology}, 12(7), 625--630.
https://doi.org/10.18178/ijiet.2022.12.7.1662

Gaffney, H., Farrington, D. P., \& Ttofi, M. M. (2019). Examining the
effectiveness of school-bullying intervention programs globally: A
meta-analysis. \emph{International Journal of Bullying Prevention},
1(1), 14--31. https://doi.org/10.1007/s42380-019-0007-4

Gaffney, H., Ttofi, M. M., \& Farrington, D. P. (2021). What works in
anti-bullying programs? Analysis of effective intervention components.
Journal of School Psychology, 85, 37--56.
https://doi.org/10.1016/j.jsp.2020.12.002

Gillespie, G. L., Brown, K., Grubb, P., Shay, A., \& Montoya, K. (2015).
Qualitative evaluation of a role play bullying simulation. Journal of
Nursing Education and Practice, 5(6).
https://doi.org/10.5430/jnep.v5n6p73

Goldberg, B., \& Robson, R. (2023, June). AI to Support Guided
Experiential Learning. In International Conference on Artificial
Intelligence in Education (pp. 103-108). Cham: Springer Nature
Switzerland.

Govindaraj, M., Asha, V., Marutheesha, H., Kumar, M. D. S., Muniprasad,
M., \& Ramesh, N. (2024, April). IntelliSecure AI-Powered Intrusion
Detection Framework. In 2024 International Conference on Inventive
Computation Technologies (ICICT) (pp. 365-370). IEEE.
https://doi.org/10.1109/ICICT60155.2024.10544435

Hawker, D. S. J., \& Boulton, M. J. (2002). Twenty years' research on
peer victimization and psychosocial maladjustment: A meta-analytic
review of cross-sectional studies. In Annual Progress in Child
Psychiatry and Child Development 2000-2001 (pp. 505--534). Routledge.
http://dx.doi.org/10.4324/9780203449523-26

Holt, M. K., Vivolo-Kantor, A. M., Polanin, J. R., Holland, K. M.,
DeGue, S., Matjasko, J. L., Wolfe, M., \& Reid, G. (2015). Bullying and
suicidal ideation and behaviors: A meta-analysis. Pediatrics, 135(2),
e496--e509. https://doi.org/10.1542/peds.2014-1864

Kolb, D. A. (2015). Experiential learning: Experience as the source of
learning and development. Pearson Education.

Konak, A., Clark, T. K., \& Nasereddin, M. (2014). Using
Kolb\textquotesingle s Experiential Learning Cycle to improve student
learning in virtual computer laboratories. Computers \& Education, 72,
11-22.

Kärnä, A., Voeten, M., Little, T. D., Poskiparta, E., Alanen, E., \&
Salmivalli, C. (2011). Going to scale: A nonrandomized nationwide trial
of the KiVa antibullying program for grades 1--9. Journal of Consulting
and Clinical Psychology, 79(6), 796--805.
https://doi.org/10.1037/a0025740

Lomas, J., Stough, C., Hansen, K., \& Downey, L. A. (2011). Brief
report: Emotional intelligence, victimisation and bullying in
adolescents. Journal of Adolescence, 35(1), 207--211.
https://doi.org/10.1016/j.adolescence.2011.03.002

Mahdi, Jelodari., Mohammad, Hossein, Amirhosseini., Andrea,
Giraldez‐Hayes. (2023). An AI powered system to enhance self-reflection
practice in coaching. Cognitive computation and systems, doi:
10.1049/ccs2.12087

Martel-Santana, A., \& Martín-del-Pozo, M. (2023). Design, development,
and evaluation of a serious game aimed at addressing bullying and
cyberbullying with primary school students. In Lecture Notes in
Educational Technology (pp. 1246--1254). Springer Nature Singapore.
http://dx.doi.org/10.1007/978-981-99-0942-1\_131

Mendoza-Pinto, R. (2023). Artificial Intelligence in the Fight Against
Bullying: Integration of ChatGPT in an Emotional Support Chatbot.
Proceedings http://ceur-ws. org ISSN, 1613, 0073.

Menesini, E., Nocentini, A., \& Palladino, B. E. (2012). Empowering
students against bullying and cyberbullying: Evaluation of an Italian
peer-led model. International Journal of Conflict and Violence (IJCV),
6(2), 313-320.

Murphy, S., \& Faulkner, D. (2011). The relationship between bullying
roles and children\textquotesingle s everyday dyadic interactions.
Social Development, 20(2), 272-293.

Nirmal, Kumar., Deepika, Singhal. (2024). Empowering Mental Well-being:
AI-guided Self-Recognition and Support. International research journal
of computer science, doi: 10.26562/irjcs.2024.v1105.06

Nocentini, A., Colasante, T., Malti, T., \& Menesini, E. (2020). In my
defence or yours: Children's guilt subtypes and bystander roles in
bullying. European Journal of Developmental Psychology, 17(6), 926--942.
https://doi.org/10.1080/17405629.2020.1725466

Olweus, D. (1997). Bully/victim problems in school: Knowledge base and
an effective intervention program. The Irish Journal of Psychology,
18(2), 170--190. https://doi.org/10.1080/03033910.1997.10558138

Olweus, D., Limber, S., \& Mihalic, S. F. (1999). Blueprints for
violence prevention, book nine: Bullying prevention program. Boulder,
CO: Center for the Study and Prevention of Violence, 12(6), 256-273.

Orrù, G., Galli, A., Gattulli, V., Gravina, M., Micheletto, M., Marrone,
S., ... \& Sansone, C. (2023). Development of Technologies for the
Detection of (Cyber) Bullying Actions: The BullyBuster Project.
Information, 14(8), 430. https://doi.org/10.3390/info14080430

Pellegrini, A. D. (2002). Bullying, victimization, and sexual harassment
during the transition to middle school. Educational Psychologist, 37(3),
151--163. https://doi.org/10.1207/s15326985ep3703\_2

Rivas, R., Shahbazi, M., Garett, R., Hristidis, V., \& Young, S. (2020).
Mental health--related behaviors and discussions among Young adults:
analysis and classification. Journal of medical internet research,
22(5), e17224. https:10.2196/17224

Ryan, W., \& Smith, J. D. (2009). Antibullying Programs in Schools: How
Effective are Evaluation Practices? Prevention Science, 10(3), 248--259.
https://doi.org/10.1007/s11121-009-0128-y

Salmivalli, C. (1999). Participant role approach to school bullying:
Implications for interventions. Journal of Adolescence, 22(4), 453--459.
https://doi.org/10.1006/jado.1999.0239

Salmivalli, C. (2010). Bullying and the peer group: A review. Aggression
and Violent Behavior, 15(2), 112--120.
https://doi.org/10.1016/j.avb.2009.08.007

Salmivalli, C., Kaukiainen, A., \& Voeten, M. (2005). Anti‐bullying
intervention: Implementation and outcome. British Journal of Educational
Psychology, 75(3), 465--487. https://doi.org/10.1348/000709905x26011

Salmivalli, C., Voeten, M., \& Poskiparta, E. (2011). Bystanders matter:
Associations between reinforcing, defending, and the frequency of
bullying behavior in classrooms. Journal of Clinical Child \&amp;
Adolescent Psychology, 40(5), 668--676.
https://doi.org/10.1080/15374416.2011.597090

Sanchez, R., Brown, E., Kocher, K., \& DeRosier, M. (2017). Improving
Children's Mental Health with a Digital Social Skills Development Game:
A Randomized Controlled Efficacy Trial of Adventures aboard the S.S.
GRIN. Games for Health Journal, 6(1), 19--27.
https://doi.org/10.1089/g4h.2015.0108

Sanoubari, E., Cardona, J. E. M., Houston, A., Young, J., \& Dautenhahn,
K. (2022). Designing an Anti-Bullying Serious Game: Insights from
Interviews with Teachers. In Serious Games (pp. 102--121). Springer
International Publishing. http://dx.doi.org/10.1007/978-3-031-15325-9\_9

Shaikh, O., Chai, V. E., Gelfand, M., Yang, D., \& Bernstein, M. S.
(2024, May 11). Rehearsal: Simulating conflict to teach conflict
resolution. Proceedings of the CHI Conference on Human Factors in
Computing Systems. http://dx.doi.org/10.1145/3613904.3642159

Sharifzadeh, N., Kharrazi, H., Nazari, E., Tabesh, H., Edalati
Khodabandeh, M., Heidari, S., \& Tara, M. (2020). Health education
serious games targeting health care providers, patients, and public
health users: Scoping review. JMIR Serious Games, 8(1), e13459.
https://doi.org/10.2196/13459

Sudyana, A., Putri, A., \& Rosmansyah, Y. (2024). Feedback system in
educational games: A systematic literature review. Jurnal Pendidikan
Indonesia, 5(7), 304--323. https://doi.org/10.59141/japendi.v5i7.3059

Tangen, J. L. (2017). Attending to nuanced emotions: Fostering
supervisees' emotional awareness and complexity. Counselor Education and
Supervision, 56(1), 65--78. https://doi.org/10.1002/ceas.12060

Thornberg, R., \& Jungert, T. (2013). Bystander behavior in bullying
situations: Basic moral sensitivity, moral disengagement and defender
self‐efficacy. Journal of Adolescence, 36(3), 475--483.
https://doi.org/10.1016/j.adolescence.2013.02.003

Thornberg, R., Wänström, L., Elmelid, R., Johansson, A., \& Mellander,
E. (2020). Standing up for the victim or supporting the bully? Bystander
responses and their associations with moral disengagement, defender
self-efficacy, and collective efficacy. Social Psychology of Education,
23(3), 563--581. https://doi.org/10.1007/s11218-020-09549-z

Ttofi, M. M., Farrington, D. P., Lösel, F., Crago, R. V., \&
Theodorakis, N. (2016). School bullying and drug use later in life: A
meta-analytic investigation. School Psychology Quarterly, 31(1), 8--27.
https://doi.org/10.1037/spq0000120

Valdebenito, S., Eisner, M., Farrington, D. P., Ttofi, M. M., \&
Sutherland, A. (2018). School‐based interventions for reducing
disciplinary school exclusion: A systematic review. Campbell Systematic
Reviews, 14(1). https://doi.org/10.4073/csr.2018.1

Walker, J., Kelty, S. F., \& Ng Tseung-Wong, C. (2024). Bystander
intervention in coercive control: Do relationship to the victim,
bystander gender, and concerns influence willingness to intervene?
Journal of Interpersonal Violence, 39(15--16), 3791--3815.
https://doi.org/10.1177/08862605241234350

Young Oh, E., Song, D., \& Hong, H. (2019a). Interactive computing
technology in anti-bullying education: The effects of conversation-bot's
role on K-12 students' attitude change toward bullying problems. Journal
of Educational Computing Research, 58(1), 200--219.
https://doi.org/10.1177/0735633119839177

Zou, W., Yang, Q., DiFranzo, D., Chen, M., Hui, W., \& Bazarova, N. N.
(2024). Social Media Co-Pilot: Designing a Chatbot with Teens and
Educators to Combat Cyberbullying. International Journal of
Child-Computer Interaction, 100680.

Zych, I., Ttofi, M. M., Llorent, V. J., Farrington, D. P., Ribeaud, D.,
\& Eisner, M. P. (2020). A longitudinal study on stability and
transitions among bullying roles. Child development, 91(2), 527-545.
\end{quote}

%% file: tables/table_01.tex
\begin{table}[H]
\centering
\caption{Participant Demographics and Session Participation}
\label{tab:participants_final}
\scalebox{0.8}{
\begin{tabularx}{1.25\textwidth}{@{} l >{\raggedright\arraybackslash}X >{\raggedright\arraybackslash}X >{\raggedright\arraybackslash}X >{\centering\arraybackslash}X >{\centering\arraybackslash}X >{\centering\arraybackslash}X @{}}
\toprule
\textbf{ID} & \textbf{Region} & \textbf{Family Income Level} & \textbf{Academic Rank} & \textbf{Participate in Session 1} & \textbf{Participate in Session 2} & \textbf{Prototype Test Version} \\
\midrule
P1  & Iran             & Mixed         & Mixed         & Yes & No  & -     \\
P2  & Chinese Mainland & Below Average & Below Average & Yes & Yes & 0.3   \\
P3  & USA              & Above Average & Above Average & Yes & Yes & 0.3   \\
P4  & Chinese Mainland & Above Average & Above Average & Yes & No  & -     \\
P5  & Chinese Mainland & Above Average & Above Average & Yes & Yes & 0.8   \\
P6  & Hong Kong        & Above Average & Above Average & Yes & Yes & 1.0   \\
P7  & Chinese Mainland & Mixed         & Above Average & Yes & Yes & 0.1   \\
P8  & Hong Kong        & Mixed         & Average       & Yes & Yes & 0.1   \\
P9  & Chinese Mainland & Above Average & Above Average & Yes & Yes & 0.1   \\
P10 & Chinese Mainland & Above Average & Above Average & Yes & No  & -     \\
P11 & Hong Kong        & Average       & Above Average & Yes & Yes & 1.0   \\
P12 & Chinese Mainland & Mixed         & Below Average & Yes & Yes & 0.5   \\
P13 & USA              & Average       & Average       & Yes & Yes & 0.3   \\
P14 & Chinese Mainland & Average       & Above Average & Yes & Yes & 0.3   \\
P15 & Chinese Mainland & Average       & Above Average & Yes & Yes & 0.5   \\
P16 & Hong Kong        & Above Average & Above Average & Yes & Yes & 0.1   \\
\bottomrule
\end{tabularx}
}
\end{table}

%% file: tables/table_02.tex
\begin{table}[H]
\centering
\caption{Statements on System Usefulness}
\label{tab:usefulness}
\begin{tabular}{p{0.9\textwidth}}
\toprule
U1: This system helps students learn how to handle bullying more quickly. \\
U2: The immersive learning approach of this system better aids students in learning anti-bullying techniques. \\
U3: This system helps students apply anti-bullying techniques in real life. \\
U4: This system makes it easier for students to remember the knowledge on how to deal with bullying. \\ 
\bottomrule
\end{tabular}
\end{table}

%% file: tables/table_03.tex
\begin{table}[H]
\centering
\caption{Statements on Engagement and Empowerment}
\label{tab:engagement}
\begin{tabular}{p{0.9\textwidth}}
\toprule
E1: This system makes the anti-bullying learning experience more enjoyable for students. \\
E2: This system encourages students to take a more proactive approach to learning anti-bullying knowledge. \\
E3: This system increases students' confidence in stopping bullying in real life. \\
E4: This system boosts students' confidence in mastering anti-bullying knowledge. \\ 
\bottomrule
\end{tabular}
\end{table}

%% file: tables/table_04.tex
\begin{table}[H]
\centering
\caption{Profiles of Roles in Bullying Scenarios}
\label{tab:roles_fixed}
\scalebox{0.8}{
\begin{tabularx}{1.25\textwidth}{l >{\raggedright\arraybackslash}X >{\raggedright\arraybackslash}X >{\raggedright\arraybackslash}X >{\raggedright\arraybackslash}X}
\toprule
\textbf{Role} & \textbf{Personal Traits} & \textbf{Social Status} & \textbf{Behavioral Patterns} & \textbf{Possible Ways to Trigger Behavioral Change} \\
\midrule
Bully     & Good at sports, High academic performance, Prominent in appearance, Strong believes in their core values & Have lots of friends, Influential in class, Take Class positions like monitor & Want to dominate in social groups, Aggressive to those who threaten their position, Conceal anxiety and insecurity & A close friend to step in and talk directly to him/her, Teacher step in and talk directly to him/her, Parents recognize the problem and change their education \\
\addlinespace 
Victim    & Poor study performance, Deficiency in appearance, Suffer from mental disorders, Self-denial & Marginalized group member, Not well socialized, Few friends & Unable to secure external support & Support from others, Prevent inferiority, Appraised by teachers, Discover self values \\
\addlinespace
Upstander & Empathetic, A strong sense of justice & Influential to other bullying participants, Bully’s moral model, Victim’s close friends & Recognize victim’s status, Diffuse the situation with jokes, Make friends with the victim & Find trustable teachers to report bullying, With efficacy to stop bullying, Understand the outcomes of bullying \\
\bottomrule
\end{tabularx}
}
\end{table}

%% file: tables/table_05.tex
\begin{table}[H]
\centering
\caption{Comparison of Animation-based vs. Conversation-based Programs}
\label{tab:comparison_fixed}
\begin{tabularx}{\textwidth}{ >{\raggedright}p{3cm} >{\raggedright\arraybackslash}X >{\raggedright\arraybackslash}X }
\toprule
\textbf{Criteria} & \textbf{Animation-based programs} & \textbf{Conversation-based programs} \\
\midrule
Suitability for Age Groups & More suitable for younger students, such as in early childhood or elementary education (P9) & More suitable for older students or situations requiring critical thinking (P9) \\
\addlinespace
Engagement & Engagement from visual appeal (P2, P16) & Engagement through active participation in dialogue and timely feedback (P2, P9) \\
\addlinespace
Knowledge Organization & Students as passive receivers of knowledge (P9) & Student-centered active learning process (P9) \\
\addlinespace
Application in Problem Solving & May not be meaningful in urgent problem-solving situations (P9) & Strategies learnt can be applied directly. (P9) \\
\bottomrule
\end{tabularx}
\end{table}